\documentclass[fleqn,10pt]{wlscirep}
\usepackage[utf8]{inputenc}
\usepackage[T1]{fontenc}
\usepackage{pifont}
\usepackage{booktabs,subcaption,amsfonts,dcolumn}

\newcommand{\cmark}{\ding{51}}%
\newcommand{\xmark}{\ding{55}}%
\usepackage{xcolor}         
\usepackage{color, colortbl}
\definecolor{brickred}{HTML}{f03b20}
\definecolor{LightBlue}{rgb}{0.8,0.9,1}

\usepackage{lineno}

\title{CAPTURE-24: A large dataset of wrist-worn activity tracker data collected in the wild for human activity recognition}

\author[1,2]{Shing Chan}
\author[1,2]{Hang Yuan}
\author[3]{Catherine Tong}
\author[1,2,4]{Aidan Acquah}
\author[1,2]{Abram Schonfeldt}
\author[5]{Jonathan Gershuny}
\author[1,2*]{Aiden Doherty}

\affil[1]{Big Data Institute, University of Oxford, Oxford, UK}
\affil[2]{Nuffield Department of Population Health, University of Oxford, Oxford, UK}
\affil[3]{Department of Computer Science, University of Oxford, Oxford, UK}
\affil[4]{Department of Engineering Science, University of Oxford, Oxford, UK}
\affil[5]{Social Research Institute, University College London, London, UK}

\affil[*]{corresponding author: Aiden Doherty (aiden.doherty@ndph.ox.ac.uk)}

\begin{abstract}
Existing activity tracker datasets for human activity recognition are typically obtained by having participants perform predefined activities in an enclosed environment under supervision. This results in small datasets with a limited number of activities and heterogeneity, lacking the mixed and nuanced movements normally found in free-living scenarios. As such, models trained on laboratory-style datasets may not generalise out of sample. To address this problem, we introduce a new dataset involving wrist-worn accelerometers, wearable cameras, and sleep diaries, enabling data collection for over 24 hours in a free-living setting. The result is CAPTURE-24, a large activity tracker dataset collected in the wild from 151 participants, amounting to 3883 hours of accelerometer data, of which 2562 hours are annotated. CAPTURE-24 is two to three orders of magnitude larger than existing publicly available datasets, which is critical to developing accurate human activity recognition models.

\end{abstract}
\begin{document}

\flushbottom
\maketitle

\thispagestyle{empty}

\section*{Background \& Summary}


With the increasing adoption of activity trackers such as Fitbit and Apple Watch, the ability to extract objective health-related behavioural insights at an unprecedented scale prompts new opportunities in medicine.
A particularly promising direction is the use of accelerometer-based activity recognition in healthcare, where it is still common to use recall diaries or time and labour-intensive methods such as clinical fitness tests. These approaches suffer from objectivity and/or scalability issues. Wrist-worn accelerometers, being low-cost, low-powered and convenient, allow us to efficiently obtain an objective and high-resolution picture of a user's daily activities, which brings new opportunities for real-time precision medicine, digital phenotyping for routine care and clinical trials~\cite{creagh2022digital, schalkamp2023wearable, gupta2023home}, and large-scale population and epidemiological studies~\cite{master2022association, willetts2018statistical,walmsley2020reallocating}. The success of these applications depends on a reliable activity recognition model, which requires a sizeable and representative labelled dataset.

However, existing open accelerometer datasets have many shortcomings due to the data collection protocol commonly employed, whereby participants are invited to an enclosed environment to engage in a set of activities pre-defined by the experimenters, often in a given sequence and under some form of supervision.
This laboratory-style setup causes the following limitations: 1) the amount of data collected is usually small as this is a labour-intensive approach; 2) it often does not allow for mixed activities; 3) even when mixed activities are allowed, the nature of the study (instruction prompting, supervised performance, enclosed environment) encourages homogeneous prototypical movement patterns as participants are subject to acquiescence bias; 4) the sequence of activities is artificial. The latter means that such data cannot be used for sequence modelling (e.g. hidden Markov models, recurrent neural networks) -- for example, if the study were conducted in a way that the sequence of activities was fixed, sequence modelling would be highly overfitted to this sequence (the transition matrix would be sparse and close to identity), showing high in-dataset test accuracy but failing to generalise outside the dataset (note that simply randomising the sequence does not fix the problem).
The end result of all these limitations is a model that seems to perform well in the study even with proper validation and testing but underperforms in the real world.

In this work, we address these issues by releasing the CAPTURE-24 dataset -- a large, in-the-wild dataset of annotated wrist-worn accelerometer readings, which was first designed to test time-use diary against device measurements\cite{gershuny2020testing}. CAPTURE-24 includes annotated accelerometer data from 151 participants, each with around 24 hours of wear time, making it several orders of magnitude larger than existing datasets (Table~\ref{tab:data_comp}). We anticipate the CAPTURE-24 dataset to be a valuable resource in wearable sensor-based human activity recognition, especially for research in data-hungry methods such as deep learning. We illustrate this in our benchmarks, which include commonly used methods such as random forest, XGBoost, hidden Markov models, and deep learning methods.

\begin{table}
    \small
    \caption{\textbf{Publicly available wrist-worn accelerometer datasets.} hrs: hours; ppl: people; mins: minutes.}    \label{tab:data_comp}
    \begin{tabular}[b]{llllll}
        \toprule
        Dataset & Size & Annotations\dag & Domain & Free-living & Year \\
        \midrule
        \rowcolor{LightBlue}
        CAPTURE-24 & 2562 hrs (151 ppl × 24 hrs)$^\ddag$ & >200 & Leisure, Sports, Occupation & \cmark & 2024 \\
        Clemson~\cite{mattfeld2017} & 120 hrs (30 ppl x 240 mins) & Steps & Leisure & \xmark & 2017 \\
        WISDM~\cite{weiss2019} & 43 hrs (51 ppl × 54 mins) & 18 & Leisure & \xmark & 2019 \\
        REALDISP~\cite{banos2012} & 39.8 hrs (17 ppl x 140.5 mins) & 33 & Sports & \xmark & 2012 \\
        OxWalk~\cite{small2023} & 39 hrs (39 ppl x 60 mins) & Steps & Leisure & \cmark & 2023 \\
        Hang-Time~\cite{hoelzemann2023} & 37.8 hrs (24 ppl x 94.4 mins) & 11 & Sports & \cmark & 2023 \\
        Leisure Activities~\cite{berlin2012} & 27.7 hrs (6 ppl x 277 mins) & 6 & Leisure & \cmark & 2012 \\
        WetLab~\cite{scholl2015} & 21.0 hrs (22 ppl x 57.2 mins) & 8 & Occupation & \xmark & 2015 \\
        Realworld~\cite{sztyler2016} & 18 hrs (15 ppl × 70 mins) & 8 & Leisure & \xmark & 2016 \\
        Smartwatch Swimming~\cite{brunner2019} & 17 hrs (40 ppl x 25.5 mins) & 5 & Sports & \xmark & 2019 \\
        WEAR~\cite{bock2023} & 15 hrs (18 ppl x 50 mins) & 18 & Sports & \cmark & 2023 \\
        TNDA-HAR~\cite{yan2023} & 5.7 hrs (23 ppl x 14.6 mins) & 8 & Leisure & \xmark & 2021 \\
        Opportunity++~\cite{ciliberto2021} & 5.3 hrs (4 ppl × 80 mins) & > 24,000 & Leisure, Gestures & \xmark & 2021 \\
        Opportunity~\cite{roggen2010} & 5.3 hrs (4 ppl × 80 mins) & 13 & Leisure, Gestures & \xmark & 2010 \\
        UC Berkley WARD~\cite{yang2009} & 5 hrs (20 ppl x 15 mins) & 13 & Leisure & \xmark & 2009 \\
        PAMAP2~\cite{reiss2012} & 4.5 hrs (9 ppl x 30 mins) & 18 & Leisure & \xmark & 2012 \\
        CMU-MMAC~\cite{frade2008} & 3.6 hrs (43 ppl × 5 mins) & 5 & Cooking & \xmark & 2008 \\
        Skoda~\cite{zappi2008} & 3 hrs (1 ppl x 180 mins) & 10 & Occupation & \xmark & 2008 \\
        ADL~\cite{bruno2013} & 2.7 hrs (16 ppl x 10 mins) & 14 & Leisure & \xmark & 2014 \\
        MHEALTH~\cite{banos2014} & 2.5 hrs (10 ppl x 15 mins) & 12 & Leisure & \xmark & 2014 \\
        UTD-MHAD~\cite{chen2015a} & 1.4 hrs (8 ppl x 10.8 mins) & 27 & Gestures & \xmark & 2015 \\
        UC Berkley HMAD~\cite{ofli2013} & 1.3 hrs (12 ppl x 6.3 mins) & 11 & Leisure, Gestures & \xmark & 2013 \\
        Daily and Sports Activities~\cite{altun2010} & 0.7 hrs (8 ppl x 5 mins) & 19 & Leisure, Sports & \xmark & 2010 \\        
        UTD-MHAD KinectV2~\cite{chen2015} & 0.4 hrs (6 ppl x 4.3 mins) & 10 & Gestures & \xmark & 2015 \\
        \bottomrule
    \end{tabular}
    \footnotesize{
    \raggedright
    \dag For datasets with annotations at different levels of granuality, the largest number of annotations at a given level of granularity was chosen.\\
    $\ddag$ We released 3883 hours of total recording, of which 2562 hours are labelled. \\}
\end{table}

\section*{Methods}



\subsection*{Data Acquisition}
\begin{figure}[t]
	\centering
	\includegraphics[width=\textwidth]{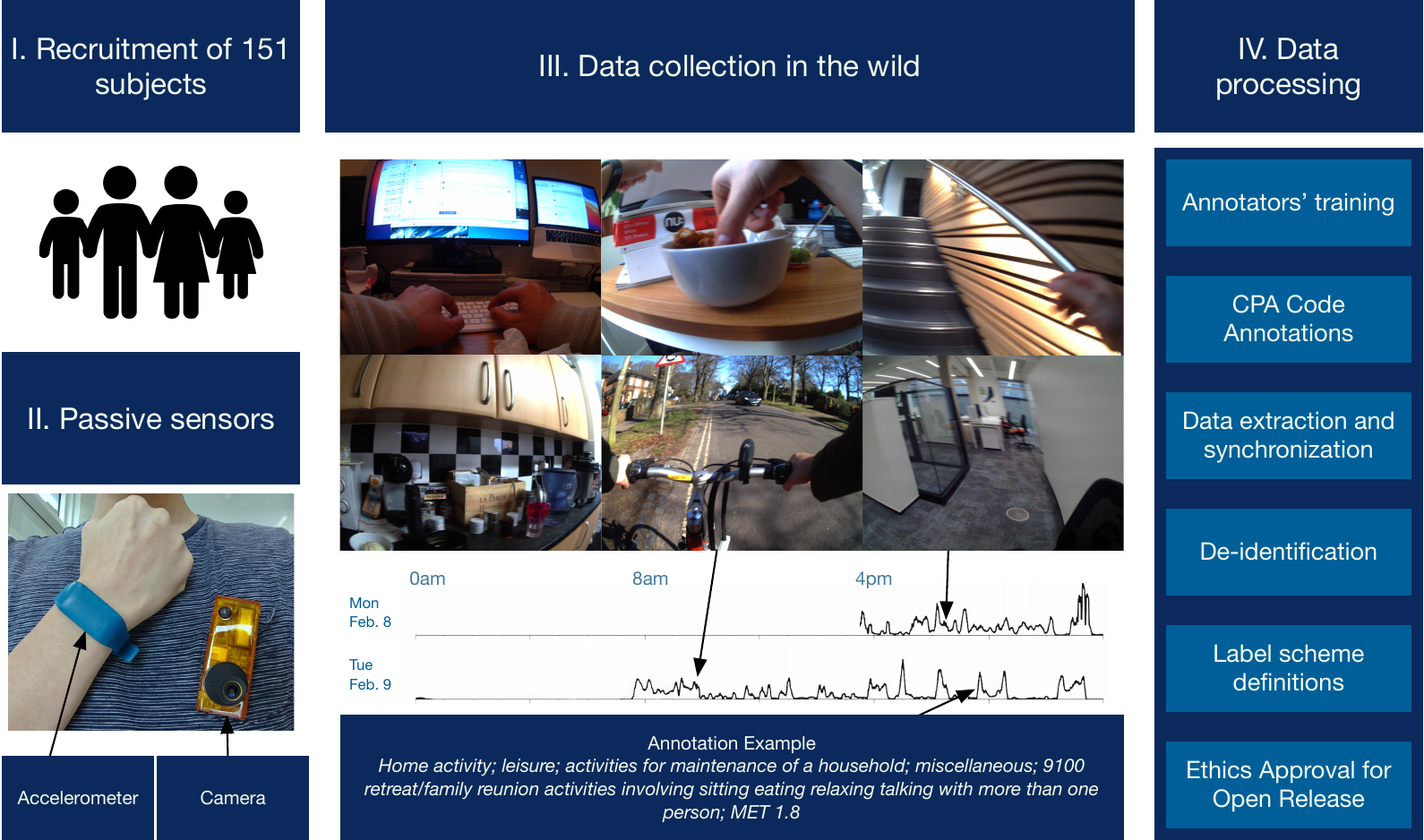}
	\caption{\textbf{Overview of the creation of the CAPTURE-24 Dataset}. Recruited subjects wore an activity tracker for roughly 24 hours. They also wore a camera during daytime and used a diary to register their sleep times during nighttime. The collected data was processed and harmonised to obtain acceleration time-series data annotated with the activities performed. CPA: Compendium of Physical Activities; MET: metabolic equivalent. Note that the camera images are not part of the dataset release.}
    \label{fig:overview}
\end{figure}

The CAPTURE-24 study was the first sizeable attempt to test traditional self-report time-use diaries against real-time passive sensing instruments, namely, wearable cameras and activity trackers~\cite{gershuny2020testing}. Data collected from this study (carried out in 2014-2015) forms the majority of our CAPTURE-24 dataset -- the data being released includes additional data collected since then. Additional processing, labelling for activity recognition, and anonymization were conducted to permit its open release. An overview of the procedures carried out is depicted in Figure~\ref{fig:overview}.

The design and associated operating procedures of the CAPTURE-24 study were based on the findings of a pilot study $(n=14)$ \cite{kelly2015developing}. Members of the public from Oxfordshire, United Kingdom, were recruited as study participants following advertisements with a £20 voucher for taking part. A member of the research team met with participants to explain the project purpose, gain written informed consent, complete a short demographic questionnaire (including gender, age, height and weight) and deliver the instruments. During the designated data collection day, participants were asked to wear a wrist-worn accelerometer continuously and a wearable camera while awake. For sleep monitoring, participants were asked to complete a simple sleep diary consisting of two questions: ``what time did you first fall asleep last night?'' and ``what time did you wake up today (eyes open, ready to get up)?''. Participants were also asked to keep a harmonised European time-use survey\cite{statistical2009harmonised}, from which sleep information was extracted when data was missing from the simple sleep diary. 
An initial 166 participants were recruited, of which 151 remained after disregarding participants with incomplete, corrupted and/or bad quality data.

\paragraph{Accelerometer}
Participants were asked to wear an Axivity AX3 wrist-worn tri-axial accelerometer on their dominant hand. The accelerometer was set to capture tri-axial acceleration data at 100 Hz with a dynamic range of $\pm$8g. Axivity device has been validated for estimating energy expenditure in a free-living environment\cite{white2019estimating}. This device has also demonstrated equivalent signal vector magnitude output on multi-axis shaking tests with other commonly-used accelerometers\cite{ladha2013shaker}.

\paragraph{Wearable Camera}
Wearable cameras were used to collect ground truths of the participants' activities while wearing the accelerometers. Participants were given an OMG Life Autographer, a wearable camera worn around the neck which automatically takes photographs every 20 - 40 seconds and has up to 16 hours battery life and storage capacity for over one week’s worth of images~\cite{doherty2013wearable}. When worn, the camera is reasonably close to the wearer’s eye line and has a wide-angle lens to CAPTURE the wearer’s view~\cite{hodges2006sensecam}. Previously, annotations from wearable cameras have been found to have strong agreement with the more expensive direct observation methods to classify activity types (inter-rater reliability via Cohen's $\kappa$ of 0.92)\cite{kelly2015developing}. Recently, the camera images have also been found to correctly identify $85\%+$ of sitting time against direct observations \cite{martinez2021validation}. Some sample images captured by the wearable camera can be seen in Figure~\ref{fig:overview}.

Due to the intrusive nature of wearable cameras, we abide by the ethical framework established by Kelly and colleagues through data collection and processing; this included scheduling a reviewing session with participants who revisited their own camera data to remove any unwanted or sensitive images\cite{kelly2013ethics}. The public CAPTURE-24 dataset also excludes image data from the wearable cameras -- only text annotations of the images are provided.

\paragraph{Data Annotation}
We relied on the time-stamped wearable camera images to annotate the accelerometer data during wake time, and sleep diaries during sleep time. To standardize the annotation taxonomy, we employed activity codes from the Compendium of Physical Activities (CPA)~\cite{ainsworth20112011}. This describes activities and their contexts in a hierarchical fashion with an associated Metabolic Equivalent of Task (MET) score to represent the mass-specific energy expenditure of activities. An example found in CAPTURE-24 is \texttt{``occupation; interruption; 11795 walking on job and carrying light objects such as boxes or pushing trolleys; MET 3.5''}. 
To ensure the reliability of the annotation process, all annotators had to complete a short training course. This covered ethics training for handling image data, usage of a specifically-developed image browsing software~\cite{doherty2011automatically}, annotation training, and finally passing annotation quality checks on a held-out gold-standard dataset, where annotators have to achieve an (Cohen's) inter-rater agreement score of $\kappa>0.8$.

\subsection*{Data Processing} \label{sec:data_process}

\paragraph{Data Extraction}
The Axivity Omgui software\footnote{https://github.com/digitalinteraction/openmovement/wiki/AX3-GUI} distributed by the accelerometer manufacturers was used for initialization of the measurements, synchronization, and downloading binary accelerometry files recorded on the devices. On top of this, we applied sampling rate correction by nearest-neighbor interpolation to fix any irregular sampling that may happen due to machine error. To correct for any accelerometer miscalibration and measurement drifts, gravity autocalibration~\cite{van2014autocalibration} was applied to reduce discrepancies across devices.

\paragraph{De-identification} To protect participant privacy, we selected a subset of the collected data for public release -- the accelerometer data, the text annotations, and the participants' gender and age. Image data is not included in the release.
Participant ages were binned into 4 similarly-sized groups $\{$``18-29'', ``30-37'', ``38-52'' and ``53 or above''$\}$.
For further de-identification, actual dates were randomized and timestamps were shifted by a small random amount.
We also reconsidered the acquired CPA code annotations containing sensitive or rare activities, judging on a case-by-case basis whether to simplify the annotation to ensure anonymization. A hypothetical example is a code containing the description \texttt{``skiing, cross country, >8.0 mph, elite skier, racing; MET 15''} which comes from only one participant in the dataset (a professional skier), the annotation will be re-labeled as \texttt{``sports; MET 15''}. The inclusion of the original label's MET ensures that the activity intensity is still retained in the annotation, but only in cases where its inclusion does not permit the activity to be uniquely identified independently from the CPA, in which case the MET value is rounded to the nearest, most commonly occuring MET value seen in CAPTURE-24.

\subsection*{Benchmarks}
We considered two activity recognition tasks: one to classify intensity levels of physical activity, and another to classify activities of daily living. For these, we re-worked the CPA codes accordingly, simplifying and mapping them to two sets of labels \cite{willetts2018statistical, walmsley2020reallocating}.
The labels for intensities of physical activity are \{``sleep'', ``sedentary'', ``light physical activity'', ``moderate-to-vigorous physical activity''\},
and the labels for activities of daily living are
\{``sleep'', ``sitting'', ``standing'', ``household-chores'', ``manual-work'', ``walking'', ``mixed-activity'', ``vehicle'', ``sports'', ``bicycling''\}.

\subsubsection*{Data preprocessing for activity recognition}

We followed the common practice of sliding-window segmentation \cite{bulling2014tutorial} to extract fixed-size, non-overlapping windows of ten seconds. This resulted in a dataset of
$n=922,199$
windows in total, each with dimension (3, 1000) (10~sec, 100~Hz, tri-axial). Data from 100 participants (\texttt{P001-P100}, with 
$618,129$
windows) were used for model derivation, while the rest (\texttt{P101-P151}, with 
$304,070$
windows) was set aside for model evaluation; we refer to these as the Derivation Set and Test Set respectively. 
The class distribution in both sets remained similar (see Figure~\ref{fig:actDist} in the Appendix). 

\subsubsection*{Models}
We consider the following commonly used methods in the activity recognition literature:

\begin{itemize}

    \item \textbf{Random forest (RF)} {} We used a balanced random forest~\cite{chen2004using} with 3000 trees. The number of trees were chosen to be as large as possible. The model was very robust to the remaining hyperparameter choices, so we used the recommended default values.
    
    \item \textbf{XGBoost} {} We used XGBoost~\cite{chen2016xgboost} and Bayesian optimization~\cite{bergstra2013making} to tune the hyperparameters (number of estimators, max depth, gamma, regularization coefficients) with 100 iterations, although we found that it did not significantly improve upon the default hyperparameters.
    
    \item \textbf{Convolutional neural network (CNN)} {} We use 1D convolutions, residual blocks~\cite{he2016identity}, and anti-aliased downsampling~\cite{zhang2019making}. We tuned the kernel sizes, number of blocks, and number of filters using grid search and the ASHA scheduler~\cite{li2018system}.
    See Appendix~\ref{app:hyper} for further details.
    
    \item \textbf{Recurrent neural network (RNN)} {} The backbone of the architecture is the CNN mentioned above, with the second last layer (originally a fully-connected layer) replaced by a bidirectional Long Short-Term Memory module~\cite{hochreiter1997long}. This model can take a sequence of windows to model temporal dependencies. A maximum sequence length of 8 (80~sec) is used.
    See Appendix~\ref{app:hyper} for further details.
    
    
    \item \textbf{Hidden Markov models (HMM)}  Additionally, we consider the application of hidden Markov models on top of all aforementioned models to model the temporal dependencies between windows. The HMM is applied post-hoc to the final sequence of outputs from the base models.     Note that while RNN is already a temporal model, we found further improvements when applying HMM on top.

    
\end{itemize}

For RF and XGBoost, we extracted time-series features from the accelerometry that are commonly used in the literature~\cite{twomey2018comprehensive} including time and frequency domain features, angular and peak features, resulting in a total of 40 features per window. See Appendix~\ref{app:feats} for the full list of features.

\subsubsection*{Metrics}
The distribution of activities was highly imbalanced reflecting that the free-living nature of the collected data (``sleeping'', ``sitting'' and ``standing'' make up more than 60\% of activities). As a result, we reported our evaluations using metrics which are more appropriate for this such as macro-averaged F1-score, Cohen's $\kappa$, and Pearson-Yule's $\phi$ coefficient~\cite{yule1912methods, cramer2016mathematical} (also known as the Matthews correlation coefficient) on the test set. we use bootstrapping ($n=100$) to estimate 95\% confidence intervals~\cite{efron1982jackknife} on all reported metrics.

\subsubsection*{Training details}
In the deep learning experiments, we further split the derivation set of 100 users into 80 users ($503,880$ windows) for training and 20 users ($125,970$ windows) for validation and early stopping. A batch size of 512 was used throughout except for the RNN model where it was reduced to 64 in response to increased computational burden due to the sequence length of 8. Stochastic gradient descent with restarts~\cite{loshchilov2016sgdr, smith2017cyclical} was used for optimization. Four data augmentation methods were explored~\cite{um2017data}: jittering, time warping, magnitude warping, and shifting. See Appendix~\ref{app:hyper} for further details.




 

\section*{Data Records}
Our dataset is hosted at the Oxford University Research Archive under the Creative Commons ``Attribution 4.0 International (CC BY 4.0)'' License, at \url{https://doi.org/10.5287/bodleian:NGx0JOMP5}. The raw accelerometry data has been processed and stored as compressed CSV files using the biobankAccelerometerAnalysis tool. For each participant, the raw  accelerometry file contains the following columns: 
\begin{itemize}
    \item Time: the timestamp for each accelerometry reading in milliseconds;
    \item X, Y, Z: the raw accelerometry along each of the axes in g;
    \item Annotation: the activity annotation using a category from the Compendium of Physical Activities.
\end{itemize}
 
In addition, we also provided the ``annotation-label-dictionary.CSV'' to provide the annotation mapping from fine-grained activity to high-level classes that be for machine learning, genetics and population health studies~\cite{willetts2018statistical,doherty2018gwas,walmsley2021reallocation}. Finally, age group and sex information for each participant are stored in ``metadata.CSV''.



\section*{Technical Validation}


\begin{table}[t]
    \small
    \begin{center}
    \caption{\textbf{Demographic information for CAPTURE-24 participants}}
    \begin{tabular}[b]{r c  c c c}
        \toprule
         & & All & Derivation Set & Test Set  \\
         & & $n$ (\%) &  $n$ (\%) &  $n$ (\%)\\
        \midrule
        \multicolumn{1}{r}{\textbf{Gender}} \\
        Male & & 52 (34.4) & 36 (36.0) & 16 (31.4) \\
        Female & & 99 (65.6) & 64 (64.0)  & 35 (68.6) \\
        & \\%
        \multicolumn{1}{r}{\textbf{Age}}   \\
        18 - 29 && 43 (28.5) & 27 (27.0) & 17 (33.3) \\
        30 - 37 && 37 (24.5) & 26 (26.0) & 14 (27.5) \\
        38 - 52 && 37 (24.5) & 24 (24.0) & 10 (19.6) \\ 
        $\geq 53$ && 34 (22.5) & 23 (23.0) & 10 (19.6) \\ 
        \bottomrule
    \end{tabular} \label{tab:demographic}
    \end{center}

\end{table}

\begin{table}[t]
    \small
    \centering
    \caption{\textbf{Annotation examples}}
    \begin{tabular}{l}
        \toprule
         Most frequent \\
         \midrule
         7030 sleeping;MET 0.95 \\
         occupation;office and administrative support;11580 office/computer work general;MET 1.5 \\
         home activity;household chores;preparing meals/cooking/washing dishes;5035 kitchen activity;MET 3.3\\
         home activity;miscellaneous;sitting;9060 sitting/lying reading without observable activities;MET 1.3\\
         occupation;office and administrative support;11580 office wok/computer work general;MET 1.5\\
         \\
         
         \toprule
         Least frequent \\
         \midrule
         home activity;household chores;5140 sweeping garage sidewalk or outside of house;MET 4.0 \\
         carrying heavy loads;MET 8.0 \\
         occupation;interruption;walking;17070 walking downstairs;MET 3.5 \\
         leisure;miscellaneous;walking;17070 descending stairs;MET 3.5 \\
         occupation;interruption;walking;17133 walking upstairs;MET 4.0 \\
         leisure;miscellaneous;21070 (generic) walking/standing combination indoor;MET 3.0 \\
         \bottomrule
    \end{tabular}

    \label{tab:top5least5anno}
\end{table}

\begin{table}[t]
    \small
    \begin{center}
    \caption{\textbf{Labels for classification tasks on CAPTURE-24 used in previous studies and their intended objectives.}}
    \begin{tabular}[b]{lll}
        \toprule
         Reference & Study objective & Labels \\
        \midrule
        
        Willetts et al.~\cite{willetts2018statistical} &  Activity recognition & bicycling, sports, vehicle, \\
          & & mixed, walking, manual-work, \\
          & & household-chores, standing, sitting, sleep \\

        Willetts et al.~\cite{willetts2018statistical} &  Activity recognition & bicycling, sit/stand, walking,  \\
         & & vehicle, mixed activity, sleep \\
      
         Doherty et al.~\cite{doherty2018gwas} & Genomic discovery & sleep, sedentary, walking, \\
         & & moderate physical activity \\
         
         Walmsley et al.~\cite{walmsley2020reallocating} & Epidemiology & sleep, sedentary behaviour, \\
         & & light physical activity, \\
         & & moderate-to-vigorous physical activity \\
         
        \bottomrule
    \end{tabular}
    \label{tab:anno_mappings}
    \end{center}
\end{table}

Initially, 166 participants were recruited. After discarding incomplete, corrupted and/or bad quality data, \textbf{151 participants} remained amounting to a total of \textbf{3883 hours} of data. The different data sources (activity tracker, camera, sleep diary) were then harmonised and processed, resulting in \textbf{2562 hours} of annotated data.
Participant demographics are summarised in Table~\ref{tab:demographic}. The majority of participants were women (66\%). Different age groups were relatively well-represented, which is important for developing models that generalize well for changes in movement patterns due to aging (e.g. walking pace).

A total of \textbf{206 unique CPA codes} were identified. The CPA codes followed a long-tail distribution (Appendix~\ref{app:anno}), dominated by the ``sleeping'' activity which constitute more than a third of activities. The most and least frequent CPA codes are shown in Table~\ref{tab:top5least5anno}.
As the 206 codes can be overly detailed, we devised six schema (included in the data release) for mapping the fine-grained codes into sets of simplified labels.

Each scheme has an intended use according to a research question. For example, for a epidemiological study focusing on physical activity levels, it may be convenient to summarise the codes into 4 classes: ``sleep'', ``sedentary activity'', ``light physical activity'', ``moderate-to-vigorous physical activity''. For a general activity recognition study, we may instead consider activities such as ``walking'', ``standing'', ``bicycling''.
Table~\ref{tab:anno_mappings} shows three schema used in previous works to answer different research questions \cite{willetts2018statistical, doherty2018gwas, walmsley2020reallocating}.

\subsection*{Benchmarks}

\begin{table}[t]
    \small
    \centering
        
    \caption{\textbf{Performance for activity recognition on the CAPTURE-24 Dataset. 95\% confidence intervals are shown in brackets.}}
       
    \begin{subtable}{\textwidth}\centering
    \caption{\small{Classifying physical activity levels}}
    \begin{tabular}[b]{llll}
        \toprule
        Model & F1-score & Cohen's $\kappa$ & Pearson-Yule's $\phi$ \\
        \midrule
        RF        & $.711 \; (.710, .713)$ & $.665 \; (.664, .667)$ & $.666 \; (.664, .668)$ \\
        XGBoost   & $.693 \; (.691, .695)$ & $.666 \; (.664, .668)$ & $.667 \; (.665, .669)$ \\
        CNN       & $.716 \; (.714, .718)$ & $.704 \; (.703, .706)$ & $.704 \; (.702, .705)$ \\
        RNN       & $.776 \; (.774, .778)$ & $.790 \; (.788, .791)$ & $.790 \; (.789, .792)$ \\
        \midrule
        RF + HMM  & $.789 \; (.788, .791)$ & $.772 \; (.771, .774)$ & $.773 \; (.771, .774)$ \\
        XGBoost + HMM &  $.783 \; (.781, .785)$ & $.772 \; (.771, .774)$ & $.773 \; (.771, .774)$ \\ 
        CNN + HMM & $.800 \; (.798, .802)$ & $.814 \; (.812, .816)$ & $.815 \; (.813, .816)$ \\
        RNN + HMM & $.787 \; (.785, .789)$ & $.800  \; (.798, .801)$ & $.801 \; (.799, .802)$ \\
        \bottomrule
    \end{tabular} \label{tab:pa_labels}
    \end{subtable}
    \vspace{1em}

    \begin{subtable}{\textwidth}\centering
    \caption{\small{Classifying activities of daily living}}
    \begin{tabular}[b]{llll}
        \toprule
        Model & F1-score & Cohen's $\kappa$ & Pearson-Yule's $\phi$ \\
        \midrule
        RF        & $.388 \; (.386, .390)$ & $.514 \; (.512, .516)$ & $.519 \; (.518, .521)$ \\
        XGBoost   & $.399 \; (.396, .401)$ & $.576 \; (.574, .577)$ & $.580 \; (.579, .582)$ \\
        CNN       & $.485 \; (.482, .488)$ & $.626 \; (.624, .627)$ & $.627 \; (.625, .629)$ \\
        RNN       & $.534 \; (.530, .536)$ & $.697 \; (.695, .698)$ & $.698 \; (.696, .700)$ \\
        \midrule
        RF + HMM  & $.486 \; (.484, .489)$ & $.637 \; (.636, .639)$ & $.641 \; (.639, .643)$ \\
        XGBoost + HMM &  $.483 \; (.480, .486)$ & $.682 \; (.680, .684)$ & $.686 \; (.684, .688)$ \\ 
        CNN + HMM & $.576 \; (.574, .580)$ & $.735 \; (.733, .736)$ & $.737 \; (.736, .739)$ \\
        RNN + HMM & $.537 \; (.535, .540)$ & $.714  \; (.713, .716)$ & $.716 \; (.715, .718)$ \\
        \bottomrule
    \end{tabular} \label{tab:adl_labels}
    \end{subtable}
    
    \label{tab:results}

\end{table}

\begin{figure}[h!]
	\centering
	\begin{subfigure}[b]{.46\linewidth}
	    \includegraphics[width=\linewidth]{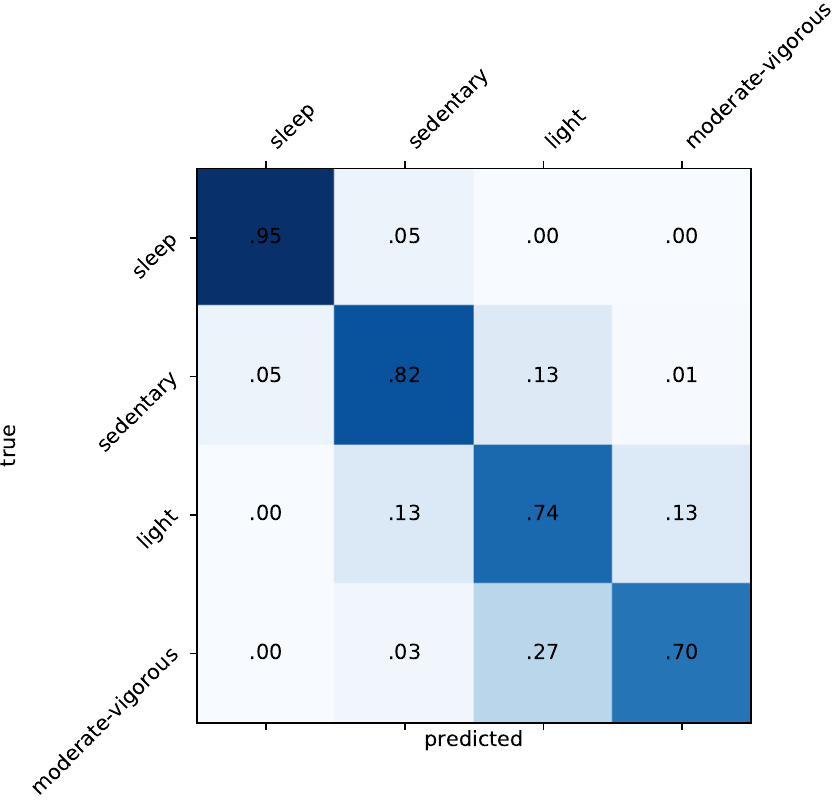}
	    \caption{\small Classifying physical activity levels}
    \end{subfigure}\hfill
    \begin{subfigure}[b]{.52\linewidth}
	    \includegraphics[width=\linewidth]{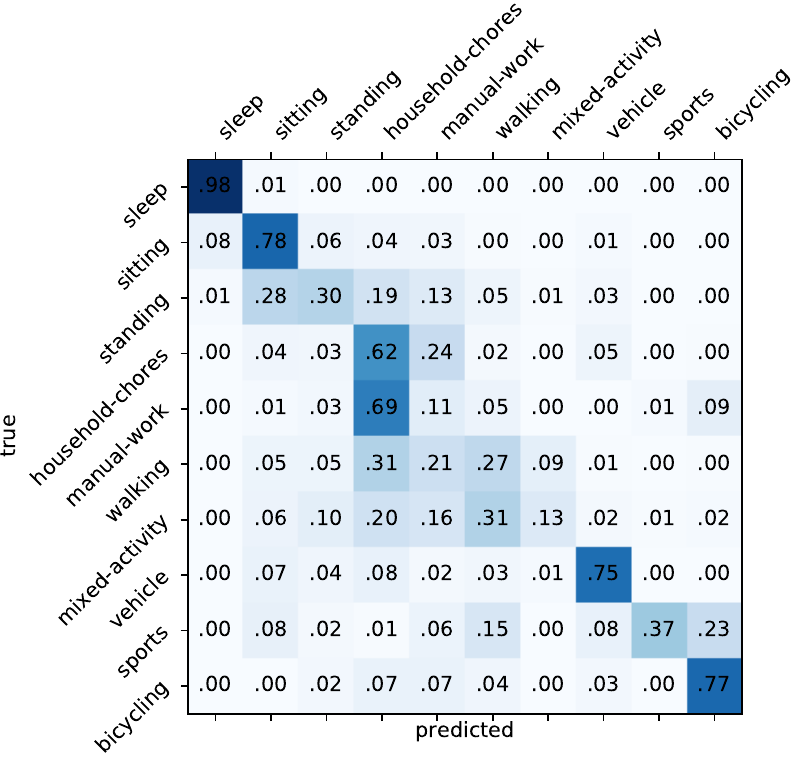}
	    \caption{\small Classifying activities of daily living}
    \end{subfigure}
    \caption{\textbf{Confusion matrix for random forest + Hidden Markov Model}}\label{fig:confusion}
\end{figure}

Results for the different models are summarised in Table~\ref{tab:results}. 
Scores for the classification of physical activity levels are shown in Table~\ref{tab:pa_labels}, and those for the classification of daily-living activities are shown in Table~\ref{tab:adl_labels}.
In each subtable, top half shows scores for the base models, and bottom half shows scores for the models enhanced with HMM smoothing. We see that HMM consistently achieves big improvements across tasks, models and metrics, highlighting the importance of modelling temporal dependencies. Further, we found that RF + HMM and XGBoost + HMM are already competitive, both performing on par or better than the more expensive models without HMM. 
The importance of temporal modelling was also seen within the models without HMM, where RNN excelled as it had the context of up to 8 consecutive windows to make predictions. Notably, we found that HMM further improved upon RNN, suggesting that longer sequence modelling would be fruitful. Interestingly, we found CNN + HMM to be the best performing model overall even though RNN performed better in the non-HMM cases.

\begin{figure}[h!]
    \centering
    \includegraphics[width=.9\linewidth]{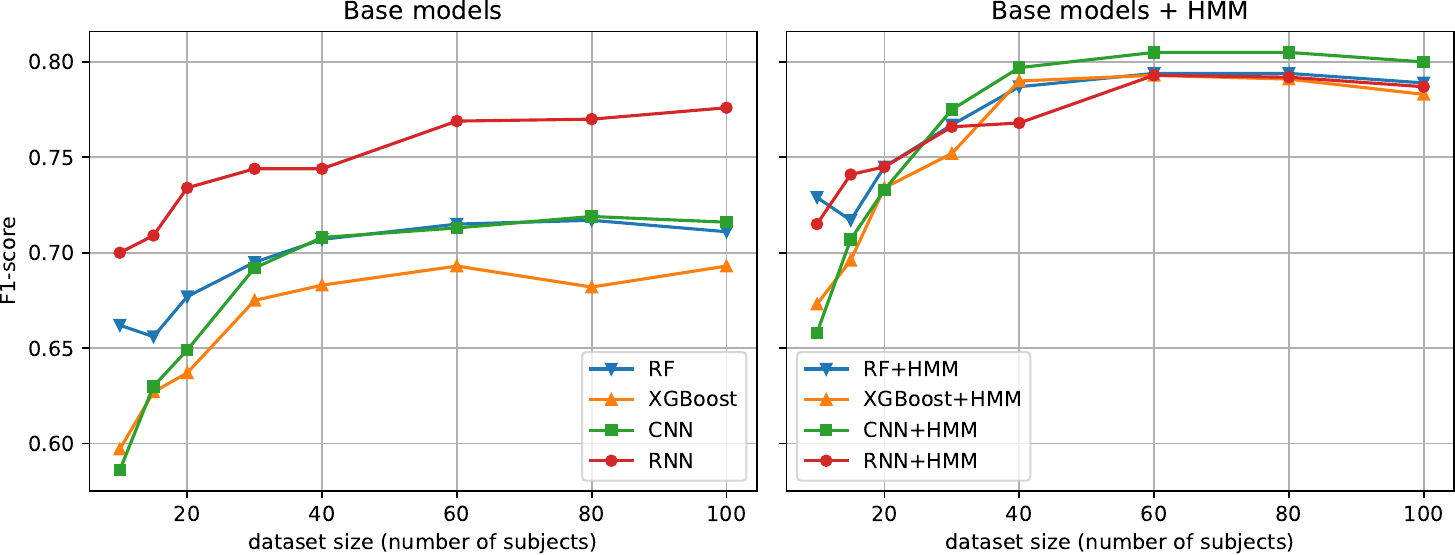}
    \caption{\textbf{F1-score as a function of dataset size for physical activity classification}}
    \label{fig:f1_vs_dsize}
\end{figure}

\paragraph{Challenges in Activity Recognition in the Wild}
We found that scores for the classification of daily-living activities were consistently lower than those for the classification of physical activity levels. More granular classification is generally harder for all types of tasks, but it is especially so with our dataset due to the ambiguity of many free-living human activities.
Figure~\ref{fig:confusion} shows the confusion matrices using the RF + HMM model. For activities of daily living, most of the confusion happens between the activities ``household-chores'', ``standing'', ``walking'', ``manual-work'' and ``mixed-activity''. This is expected given that, in free-living settings, these activities are naturally intertwined (e.g. the household chore ``cleaning, sweeping carpet or floor'' inevitably involves some degree of ``walking'' and ``standing''), as opposed to data collected in laboratory settings where the scripted activities tend to be clearly segmented and the movement patterns show less heterogeneity.
Regarding classification of physical activity intensity levels, their definitions from real-world human activities tend to be less ambiguous, therefore we observed less confusion for this classification task.

\paragraph{Performance against Dataset Size}
We highlighted the importance of having large datasets for data-intensive deep learning methods. We assessed the performance as a function of dataset size by running the benchmarks on varying number of subjects included in the derivation set (the test set of 51 subjects is unchanged). From Figure~\ref{fig:f1_vs_dsize}, we observed that in the small-data regime the outperformance of deep learning models is not so clear. In particular, if we consider only non-temporal models (RF, XGBoost and CNN), we could see that CNN only starts to outperform after around 40 subjects ($\approx 650$ person-hours). Similarly for the temporal models (RNN and $\ast$-HMM), the clear advantage of CNN-HMM became apparent only after around 30 subjects ($\approx 500$ person-hours).

\section*{Usage Notes}



We presented the CAPTURE-24 dataset to address shortcomings of existing activity recognition datasets -- namely, limited dataset sizes and unrepresentativeness due to intrusive data collection methods resulting in short time spans, limited and scripted activities, low pattern heterogeneity, and manufactured activity sequences.
We described in detail how CAPTURE-24 addressed these issues with a novel collection protocol involving indirect measurements using wearable cameras and sleep diaries, allowing for long time spans (24 hours or more) in real-world settings while also being less labour-intensive and more scalable. 
We also described procedures taken to comply with privacy and ethics standards to permit the public release of the dataset. With 2562 hours of annotated data (and 3883 hours overall), the released dataset is 2 to 3 orders of magnitude larger than existing public accelerometer datasets.

We presented benchmarks for activity recognition on this dataset with commonly used methods in the literature and discussed challenges for activity recognition in the wild. In particular, we highlighted challenges in activity recognition in the wild as many activities in the real world are intertwined, in contrast to those collected in laboratory settings.
We also highlighted the importance of having large HAR datasets for deep learning research, suggesting that existing dataset sizes are insufficient to achieve the full potential of their methods, rendering any model comparison unreliable.

\paragraph{Limitations.}
The CAPTURE-24 only contains a convenience sample of participants in Oxford. Therefore, larger datasets using more diverse populations are needed. For example, a similar dataset was collected in China for human activity recognition as part of the China Kadoorie Biobank wearable study\cite{chen2023device}. As wearable sensing technologies improve, multi-modal monitoring for human activity recognition over time has become feasible, improving the predictive power for labour-intensive activity with little wrist movement and the classification of the sleep stages.

Furthermore, Camera data may sometimes be uninformative for annotation due to obstruction, poor lighting conditions and blurriness. Since the cameras record data at a frame rate ($\approx 0.03$Hz) -- much lower than that of the accelerometers (100Hz) -- activities could have been missed. As a result, it is possible that the annotators may assign CPA codes through guess work despite our best efforts in covering uncertain scenarios in the annotator training.
A further limitation with CPA codes is that they were originally developed for use in epidemiological studies to standardise the assignment of MET values in physical activity questionnaires, thus some codes place more emphasis on distinguishing energy intensities rather than behaviours. This results in some CPA codes being ambiguous for retrospective interpreting and re-labelling. For example, the code \texttt{``home activity; miscellaneous; standing; 9050 standing talking in person/ on the phone/ computer (skype chatting) or using a mobile phone/ smartphone/ tablet; MET 1.8''} precludes distinguishing whether the participant was speaking to someone in person or through a specific device, which might have been useful in studies looking to understand people's screen-time or social behaviours. Our existing benchmark incorporates common methods used for HAR, future work could also benefit from leveraging more recent modeling techniques using DeepConvLSTM~\cite{ordonez2016deep,yuan2023self}, transformers~\cite{haresamudram2020masked}, and self-supervised learning~\cite{saeed2019multi, haresamudram2022assessing, jain2022collossl, yuan2022self}.

\paragraph{Research directions.} 

Although the feasibility of activity recognition solely from accelerometer data has been debated in recent work~\cite{tong2020accelerometers}, a proper investigation has been lacking due to the lack of realistic datasets. As mentioned, current datasets have limited and homogeneous activities which optimistically bias any assessment. The CAPTURE-24 could be useful for such investigation.
The problem of open set recognition is another interesting direction in HAR as there are practically infinite number of activities that one could consider. The fine-grained and hierarchical annotations in CAPTURE-24 can be leveraged to study this problem within the framework of zero-shot and few-shot learning.
Finally, we saw that temporal modelling using hidden Markov models consistently improved performance, but other time-series methods~\cite{ordonez2016deep, haresamudram2020masked} could be investigated leveraging the unique long-span aspect of our dataset.

\section*{Code availability}
All the code associated with this project can be used free of charge for non-commercial purposes. The annotation tool used can be accessed via the Oxford Wearable Camera Browser: \url{https://github.com/OxWearables/oxford-wearable-\\camera-browser}. The data analytics script can be accessed via biobankAccelerometerAnalysis repository:  \url{https://github.com/OxWearables/biobankAccelerometerAnalysis}. Finally, the machine learning benchmarks can be accessed via: \url{https://github.com/activityMonitoring/capture24}.



\section*{Acknowledgements} 
This work is supported by Novo Nordisk (HY, SC, AD); the Wellcome Trust [223100/Z/21/Z] (AD); GlaxoSmithKline (AA); the British Heart Foundation Centre of Research Excellence [RE/18/3/34214] (AD); the National Institute for Health Research (NIHR) Oxford Biomedical Research Centre (AD); and Health Data Research UK (RW, AD), an initiative funded by UK Research and Innovation, Department of Health and Social Care (England) and the devolved administrations, and leading medical research charities. It is also supported by the UK’s Engineering and Physical Sciences Research Council (EPSRC) with grants EP/S001530/1 (the MOA project) and EP/R018677/1 (the OPERA project) and the European Research Council (ERC) via the REDIAL project (Grant Agreement ID: 805194), and industrial funding from Samsung AI. The data collection was carried out using funding (JG) from the UK Economic and Social Research Council (grant number ES/L011662/1) and the European Research Council Advanced Grant (Grant number 339703). Finally, we would like to thank Rosemary Walmsley for her contributions in the dataset curation and analysis.
 
For the purpose of open access, the author has applied a CC-BY public copyright licence to any author accepted manuscript version arising from this submission.

\section*{Author contributions statement}
SC, HY, CT, and AD conceived the experiments. SC, HY, CT, AD, AS conducted the experiments. SC, HY, CT, AD, AA, and AS analysed the results. HY, SC wrote the first draft of the manuscript. All authors reviewed the manuscript. 

\section*{Competing interests} 
The authors declare no competing interests.

\clearpage

\bibliography{main}







\clearpage

\appendix
\section{List of hand-crafted features}\label{app:feats}
The following commonly used features~\cite{twomey2018comprehensive} (40 in total) are extracted from the raw accelerometry for the random forest and XGBoost models:

\begin{itemize}
    \item  \textbf{Quantiles} Minimum, maximum, median, 25\textsuperscript{th} and 75\textsuperscript{th} percentiles of acceleration for each of the three axis streams as well as the magnitude stream.
    \item \textbf{Correlations} Correlation between axes and 1-sec-lag autocorrelation of the magnitude stream.
    \item \textbf{Spectral features} First and second dominant frequencies and their powers, and spectral entropy.
    \item \textbf{Peak characteristics} Number of peaks and median prominence of the peaks.
    \item \textbf{Angular features} Estimated dynamic roll, pitch and yaw (mean and standard deviation), and gravity roll, pitch and yaw (mean).
\end{itemize}

\section{Hyperparameter tuning details}\label{app:hyper}

\subsection{Baseline architecture}
The final architecture is described in Table~\ref{tab:arch}. Here, $\mathrm{Conv}(k,n)$ means a 1D convolution with $n$ filters of kernel size $k$, $m \times \mathrm{ResBlock}(k,n)$ means $m$ residual blocks of size $m$ with $n$ filters and kernel size $k$~\cite{he2016identity}, $\mathrm{Drop}(p)$ is dropout~\cite{srivastava2014dropout} with rate $p$, $\mathrm{FC}(n)$ is a fully connected layer with output size $n$, $\mathrm{BiLSTM}(n)$ is a bidirectional LSTM~\cite{hochreiter1997long} with output size $n$, and finally, $\mathrm{Linear}(n)$ is a linear layer with output size $n$. As usual, batch normalization~\cite{ioffe2015batch} and rectified linear units~\cite{fukushima1982neocognitron} follow the $\mathrm{Conv}$ layers. Rectified linear units also follow the $\mathrm{FC}$ layer.
All convolutions use a stride and circular padding of $1$. Downsampling is performed with anti-aliasing as described in~\cite{zhang2019making}.

\begin{table}[ht]
    \small
    \begin{center}
    \caption{\textbf{Network architectures for convolution neural network (CNN) and recurrent neural network (RNN)}}
    \begin{tabular}[b]{ll}
        \toprule
        State size     & Layer \\
        \midrule
        (*, 3, 1000)   & Conv(3, 128) / 2 \\
        (*, 128, 500)  & Conv(3, 128), 3 x ResBlock(3, 128) / 2 \\
        (*, 128, 250)  & Conv(3, 256), 3 x ResBlock(3, 256) / 2 \\
        (*, 256, 125)  & Conv(3, 256), 3 x  ResBlock(3, 256) / 5 \\
        (*, 256, 25)   & Conv(3, 512), 3 x ResBlock(3, 512) / 5 \\
        (*, 512, 5)    & Conv(3, 512), 3 x ResBlock(3, 512) / 5 \\
        (*, 512, 1)    & Drop(0.5), FC(1024) or BiLSTM(512 (x2)) \\
        (*, 1024)      & Linear(6)  \\
        \bottomrule
    \end{tabular}    \label{tab:arch}
    \end{center}

\end{table}

For the RNN model, $\mathrm{BiLSTM}$ is used in place of $\mathrm{FC}$ in order to ingest sequences of windows -- we limit the maximum sequence length to 8. 

We tried $k\in\{3, 5\}$ for the kernel sizes and $m\in\{0,1,2,3\}$ for number of residual blocks (constrained to be the same throughout), an initial configuration of filters $n = 64 \to 64 \to 128 \to 128 \to 256 \to 256 \to 512$, and a wider one $n = 128 \to 128 \to 256 \to 256 \to 512 \to 512 \to 1024$. We used ASHA~\cite{li2018system} as implemented in Ray Tune~\cite{liaw2018tune}.

\subsection{Data augmentation}\label{app:augment}
We tried four data augmentation techniques~\cite{um2017data}: jittering, time warping, magnitude warping, and shifting. For jittering, we tried standard deviation $\sigma\in\{0, .01, .05, .1\}$. For time and magnitude warping, $\sigma\in\{0, .01, .05, .1\}$ and $\mathrm{knots}\in\{2, 4\}$. For shifting, $\mathrm{shift}\in\{0~\mathrm{sec}, 1~\mathrm{sec}, 2~\mathrm{sec}, 5~\mathrm{sec}\}$.
To reduce computational cost, each augmentation technique is tried independently and the best parameters are then combined.
Each trial is run until early-stopped with patience of 5. Table~\ref{tab:augment} reports the best parameters found. Note in particular that we did not find jittering to improve performance. For the other techniques, we found slight to moderate improvements.

\begin{table}
    \small
    \begin{center}
    \caption{\textbf{Data augmentation parameters}}
    \begin{tabular}[b]{ll}
        \toprule
        Technique & Parameters \\
        \midrule
        Jittering & $\sigma=0$ \\
        Time warping & $\sigma=.05, \; \mathrm{knots}=4$ \\
        Magnitude warping & $\sigma=.05, \; \mathrm{knots}=2$ \\
        Shifting & $\mathrm{shift}=2~\mathrm{sec}$ \\
        \bottomrule
    \end{tabular}    \label{tab:augment}
    \end{center}

\end{table}

\subsection{Optimization}
Initially, we tuned the architecture and data augmentation parameters using Adam~\cite{kingma2014adam} with learning rate $\eta=3\times10^{-3}$. After tuning was done, we retrained the optimal model using stochastic gradient descent with restarts~\cite{loshchilov2016sgdr, smith2017cyclical} and tried initial learning rates $\eta\in \{.1, .15, .2, .25, .3, .35, .4, .45\}$. 
The trials were run until early-stopped with patience of 5. The CNN model converged in around 30 epochs while the RNN model in around 40 epochs.

\subsection{Computational resources}
Models were trained using a V100 GPU with 32GB of RAM. Training time varies by model and task, but can all be completed within 12h.

\section{Annotations} \label{app:anno}
\begin{figure}[t]
	\centering
	\includegraphics[width=\textwidth]{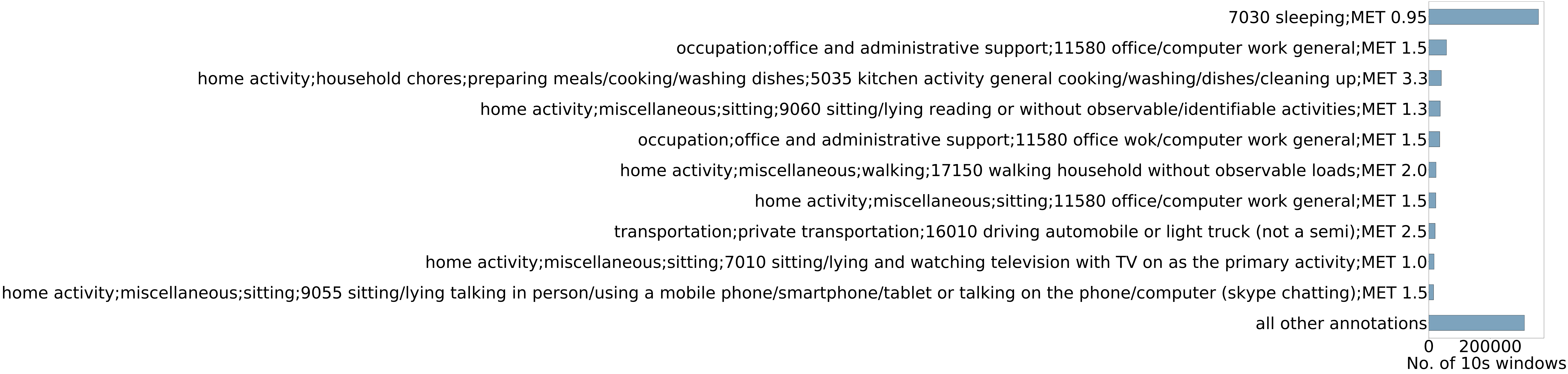}
	\caption{\textbf{Top 10 most frequent Compendium of Physical Activities code annotations found in Capture-24}}
    \label{fig:anno_dist}
\end{figure}

Figure~\ref{fig:anno_dist} plots the occurrence of the 10 most common CPA code annotations found in Capture-24. The rest of the annotations are displayed as ``all other annotations'' in the diagram to indicate the long-tail distribution over the codes. 


	

\section{Other datasets} \label{app:others}
Scores for other public datasets using the same benchmark models are shown in Table~\ref{tab:others}. As these datasets are very small, the test scores can have high variance as well as being prone to $p$-hacking (it is tempting to redo the train/test split several times to get a desired conclusion). We therefore perform leave-one-subject-out cross-testing. For WISDM~\cite{weiss2019smartphone} dataset (51 subjects), 10-fold cross-testing is used instead. For simplicity, we show results for RF and CNN only, each being an archetype of traditional and modern methods, respectively.
Unsurprisingly, we observe that CNN underperforms in the smaller datasets (ADL~\cite{bruno2013analysis} and PAMAP2~\cite{pamap2}) while RF is rather consistent across dataset sizes. On the other hand, CNN performs on par or better than RF in the larger datasets (RealWorld~\cite{realworld} and WISDM), as well as in CAPTURE-24 (results in the main text).
We also note that the performances are overall higher than those of CAPTURE-24, which is expected as these datasets are collected in a clean lab setting.

\begin{table}
    \small
    \caption{\textbf{
    Scores (median and interquartile range) for other public datasets using same benchmark models. 
    }}
    \centering
    \begin{tabular}[b]{llll} 
    
        \multicolumn{4}{c}{ADL ($n=2.7$~hrs)} \\
        \toprule
        Model & F1-score & Cohen's $\kappa$ & Pearson-Yule's $\phi$ \\
        \midrule
        RF        & $.777 \; (.671, .868)$ & $.694 \; (.642, .846)$ & $.734 \; (.661, .857)$ \\
        CNN       & $.604 \; (.568, .697)$ & $.558 \; (.484, .646)$ & $.576 \; (.538, .679)$ \\
        \bottomrule
        \\
    
        \multicolumn{4}{c}{PAMAP2 ($n=4.5$~hrs)} \\
        \toprule
        Model & F1-score & Cohen's $\kappa$ & Pearson-Yule's $\phi$ \\
        \midrule
        RF        & $.810 \; (.762, .829)$ & $.810 \; (.751, .824)$ & $.812 \; (.756, .826)$ \\
        CNN       & $.685 \; (.625, .708)$ & $.696 \; (.619, .710)$ & $.711 \; (.629, .721)$ \\
        \bottomrule
        \\

        \multicolumn{4}{c}{RealWorld ($n=18$~hrs)} \\
        \toprule
        Model & F1-score & Cohen's $\kappa$ & Pearson-Yule's $\phi$ \\
        \midrule
        RF        & $.775 \; (.674, .857)$ & $.732 \; (.620, .824)$ & $.743 \; (.640, .827)$ \\
        CNN       & $.806 \; (.700, .884)$ & $.771 \; (.639, .871)$ & $.781 \; (.654, .876)$ \\
        \bottomrule
        \\
        
        \multicolumn{4}{c}{WISDM ($n=43$~hrs)} \\
        \toprule
        Model & F1-score & Cohen's $\kappa$ & Pearson-Yule's $\phi$ \\
        \midrule
        RF        & $.805 \; (.747, .824)$ & $.752 \; (.689, .776)$ & $.756 \; (.693, .778)$ \\
        CNN       & $.804 \; (.755, .846)$ & $.747 \; (.714, .804)$ & $.752 \; (.727, .807)$ \\
        \bottomrule
        
    \end{tabular} \label{tab:others}
    
\end{table}

\section{Distribution of Coarse Activity Labels} \label{app:actdist}
\begin{figure}[t]
	\centering
	\begin{subfigure}{.75\textwidth}
	    \includegraphics[width=\textwidth]{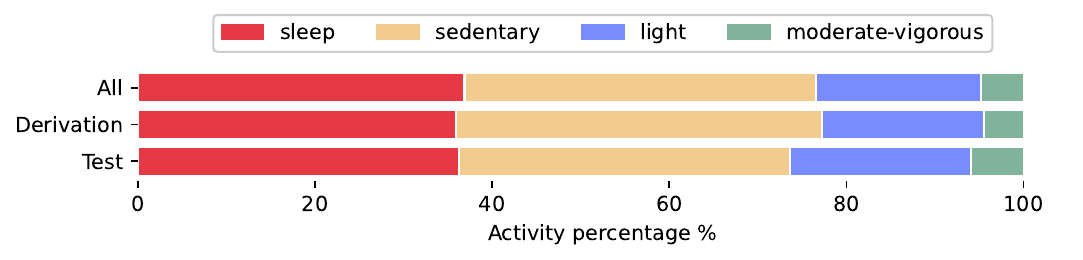}
	    \caption{\small Four classes}
	\end{subfigure}
	
	\begin{subfigure}{.75\textwidth}
		\includegraphics[width=\textwidth]{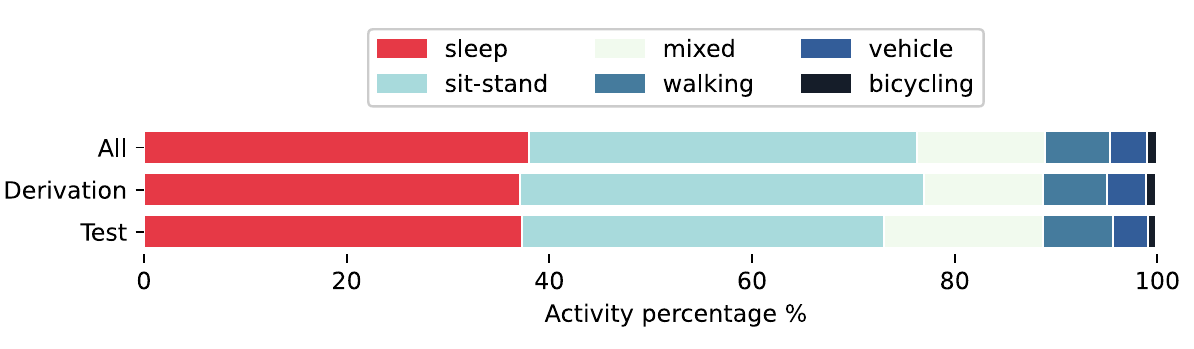}
	\caption{\small Six classes}
    \end{subfigure}
	
	\begin{subfigure}{.75\textwidth}
	    \includegraphics[width=\textwidth]{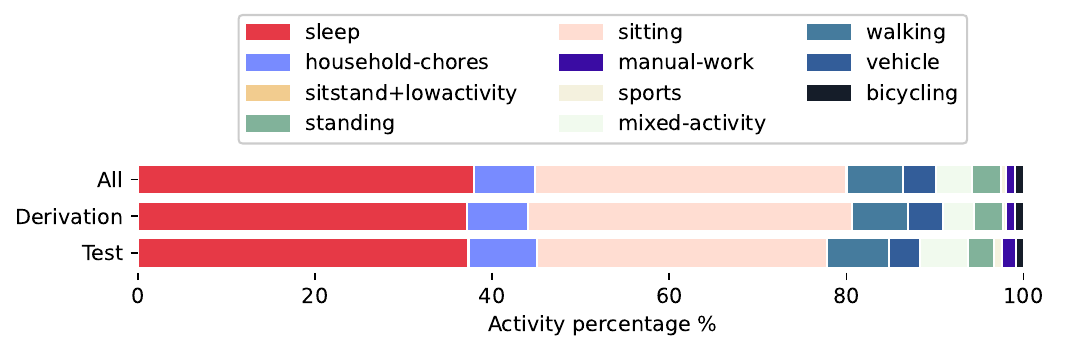}
	    \caption{\small Ten classes}
	\end{subfigure}

    \caption{\textbf {Distribution of activities different labelling schema}}
    \label{fig:actDist}
\end{figure}

In Figure~\ref{fig:actDist}, we show the activity distribution using the 6-class and 10-class schemes\cite{willetts2018statistical}.

\end{document}